\documentclass[
reprint,noeprint,
 amsmath,amssymb,
 aps,prl
]{revtex4-1}

\usepackage[dvipdfmx,dvipdfmx]{graphicx}
\usepackage{graphicx}
\usepackage{color}
\usepackage{hyperref}
 
\usepackage{subfigure} 

\usepackage{epsfig}
\usepackage{float}
\usepackage{epstopdf}
\usepackage{mathtools}
\DeclarePairedDelimiter\bra{\langle}{\rvert}
\DeclarePairedDelimiter\ket{\lvert}{\rangle}

\DeclarePairedDelimiterX\braket[2]{\langle}{\rangle}{#1 \delimsize\vert #2}

\begin{document}

\title{Deterministic generation of a four-component optical cat state
}

%\date{\today}

\author{Jacob Hastrup, Jonas S. Neergaard-Nielsen and Ulrik L. Andersen}
\affiliation{
 Center for Macroscopic Quantum States (bigQ), Department of Physics, Technical University of Denmark, Building 307, Fysikvej, 2800 Kgs.~Lyngby, Denmark}

\begin{abstract}
The four-component cat state represents a particularly useful quantum state for realizing fault-tolerant continuous variable quantum computing. While such encoding has been experimentally generated and employed in the microwave regime, the states have not yet been produced in the optical regime. Here we propose a simple linear optical circuit combined with photon counters for the generation of such optical four-component cat states. This work might pave the way for the first experimental generation of fault-tolerant optical continuous variable quantum codes. 
\end{abstract}

\maketitle

Quantum continuous variables (CV) have recently emerged as a promising platform for scalable quantum computing and communication. The main challenge - as for any other quantum information platform - is the ability to manipulate, store and communicate CV quantum information in a fault-tolerant manner in the presence of noise. In order to cope with noise, different bosonic error correction codes have been proposed, including the Gottesman-Kitaev-Preskill (GKP) codes, the cat codes and the binomial codes \cite{Gottesman2001,Cochrane1999,leghtas2013hardware,Li2017,Michael2016,Albert2018,grimsmo2019quantum}. These codes have recently been experimentally generated in microwave cavity fields coupled to superconducting circuits~\cite{Ofek2016a,hu2019,Campagne-Ibarcq2019} and in the vibrational mode of a single trapped ion~\cite{Fluhmann2018,Fluhmann2019}, and have been used to demonstrate quantum error correction and universal gate set operations.  

While the superconducting circuit and ion platforms are highly suitable for the storage and manipulation of quantum information, they are less suitable for communication over larger distances. Bosonic error-correcting codes for long-distance communication will eventually require the usage of a low-loss optical platform where the codes will be embedded in the CV optical quadratures of light~\cite{Weedbrook2012,muralidharan2016optimal}. Moreover, optical encoding is not only relevant for communication: There is an increasing interest in CV optical quantum computing partly fuelled by the recent advances in producing 1D~\cite{pysher2011parallel,Yokoyama2013} and 2D~\cite{Asavanant2019,Larsen2019} cluster states of continuous variables. 

There have been several theoretical proposals on the generation of optical GKP codes using either deterministic or probabilistic schemes. The most feasible approach is based on linear optics and photon counting detectors in which the required, and notoriously difficult, optical non-linear transformation is enabled by the non-Gaussian photon counter~\cite{Eaton2019,Su2019}. Another interesting approach requires an initial resource of cat states from which GKP states can be grown with a linear optical beam splitter network and homodyne detection \cite{vasconcelos2010all,Weigand2018}.

On the other hand, there are very few proposals for the direct generation of cat codes in the optical regime and the hope is that this might be significantly simpler than the generation of GKP states. Cat codes \cite{leghtas2013hardware,Li2017} consist of four-component cat states comprising superpositions of four coherent states, in contrast to the more common optical cat state which is a superposition of two coherent states. These latter states have undergone numerous experimental studies and have been produced in the optical regime using probabilistic approaches based on linear optics and photon counting~\cite{Dakna1997,Ourjoumtsev2006,Neergaard-Nielsen2006,Wakui2007,Gerrits2010}, and very recently, using a deterministic approach based on the Jaynes-Cumming interaction between light and a single atom in a high-finesse cavity~\cite{Hacker2019}.

One approach for generating four-component cat states was very recently proposed by Thekkadath et al. \cite{thekkadath2019engineering}. Their method uses photon number resolving detectors and coherent state ancillas to project one mode of a two-mode squeezed vacuum state into an approximate two- or four-component cat state. However, their method is probabilistic, with low success probability for larger cat states, and furthermore it requires high two-mode squeezing to obtain four-component cat states with high fidelity.

In this article, we propose a simple circuit for the deterministic generation of an optical four-component cat state based on linear optics and photon counting using an initial resource of either two-component cat states or single-photon-subtracted squeezed states. While using two-component cat states will produce exact four-component cat states, the usage of single-photon-subtracted squeezed states is able to produce approximate four-component cat states with reasonable amplitudes.

\begin{figure}
    \centering
    \includegraphics{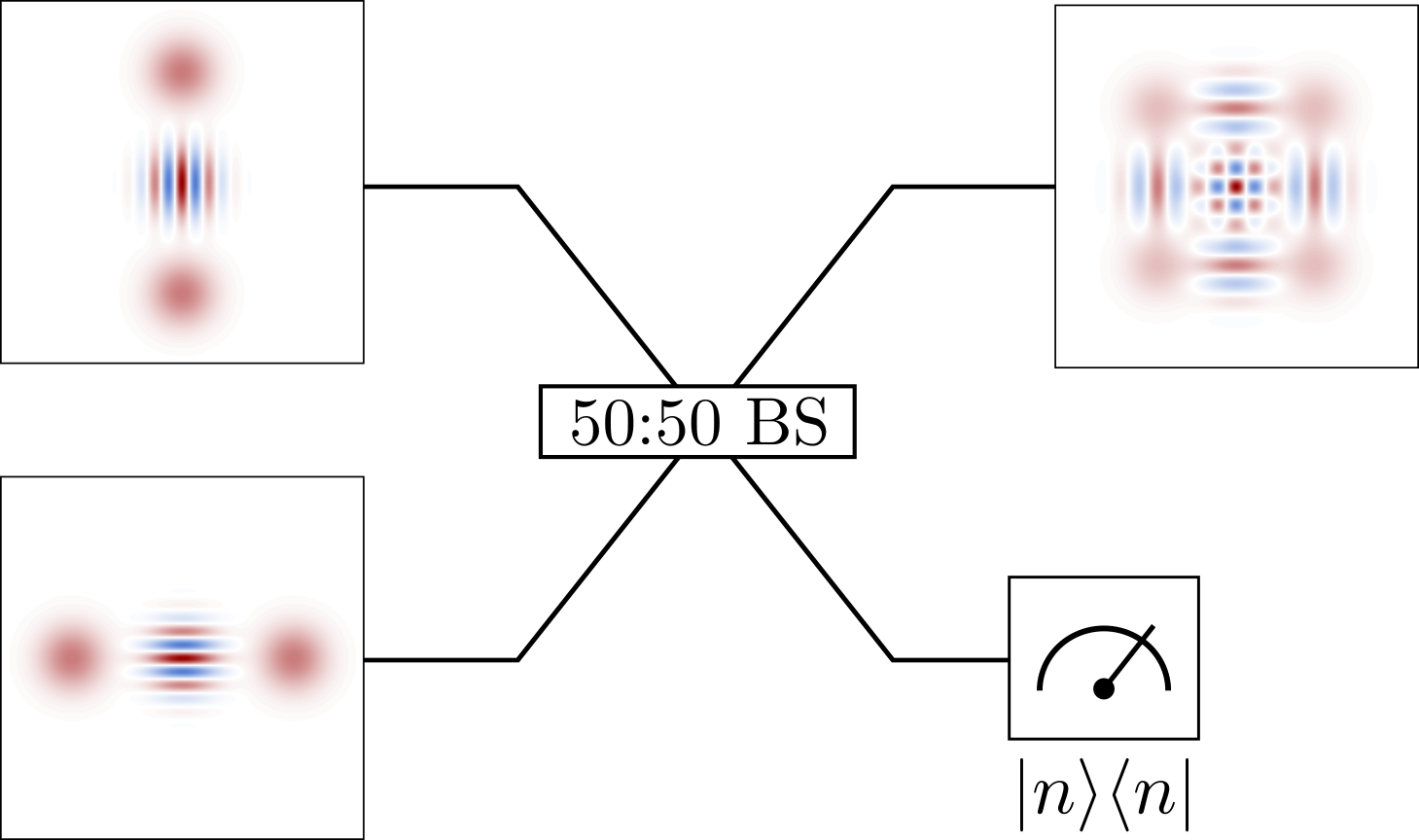}
    \caption{Schematic of proposed idea for generating four-component cat states using a balanced beam splitter and a projecting measurement. The state is generated probabilistically by projecting onto vacuum with an on-off photon counter (or a heterodyne detector) or is generated deterministically by projecting onto any photon number state using a photon number resolving detector.}
    \label{fig:Scheme}
\end{figure}

One can define four mutually orthogonal four-component cat states as

\begin{align}
|\Phi_k\rangle&=\frac{1}{N_k}\left(\ket{\beta}+(-1)^k\ket{-\beta}+(-i)^k\ket{i\beta}+i^k\ket{-i\beta}\right) \nonumber\\
&\propto\sum_{n=0}^{\infty}\frac{\beta^n}{\sqrt{n!}}\delta_{n \textrm{(mod 4)},k}\ket{n}
\label{4cat}
\end{align}
for $k=0,1,2,3$, where $N_k$ is a normalization factor and $\beta$ is the coherent state amplitude. $\delta_{a,b}$ is the Kronecker delta, i.e. the 4-component cat states have support on every 4th photon number state. The main result of this article is that these states can be readily produced by interfering two two-component cat states on a balanced beam splitter followed by a projective measurement as illustrated in Fig.~\ref{fig:Scheme}. If the two input cat states are given by  $(\ket{\alpha}+\ket{-\alpha})/N_\alpha$ and $(\ket{i\alpha}+\ket{-i\alpha})/N_\alpha$ respectively, where $\alpha$ is the cat state amplitude and $N_\alpha=(2(1+e^{-2|\alpha|^2}))^{1/2}$ is the two-component normalization factor, the beam splitter $\hat{U}_\textrm{BS}=e^{\pi/4\left(\hat{a}_1^\dagger\hat{a}_2 - \hat{a}_1\hat{a}_2^\dagger\right)}$ transforms the input state as 
\begin{eqnarray}
(\ket{\alpha}_1+\ket{-\alpha}_1)(\ket{i\alpha}_2+\ket{-i\alpha}_2)\xrightarrow{50:50 \textrm{ BS}}\nonumber \\ \ket{\beta}_1\ket{i\beta}_2+\ket{-i\beta}_1\ket{-\beta}_2
+\ket{i\beta}_1\ket{\beta}_2+\ket{-\beta}_1\ket{-i\beta}_2\nonumber
\end{eqnarray}
where $\beta=\alpha e^{i\pi/4}$. By transforming mode 2 into the Fock basis, the output state, $\ket{\Psi}$, can be written as
\begin{eqnarray}
\ket{\Psi}=\frac{e^{-|\beta|^2/2}}{N_\alpha^2}\sum_{n=0}^\infty \frac{(i\beta)^n}{\sqrt{n!}}\big(\ket{\beta}_1+(-1)^n\ket{-\beta}_1\nonumber\\
+(-i)^n\ket{i\beta}_1+i^n\ket{-i\beta}_1\big)\ket{n}_2,
\end{eqnarray}
It is clear that by projecting mode 2 onto a photon number state $\ket{n}_2$, using a photon number resolving detector (PNRD), the resulting state in mode 1 is the exact four-component cat state given in Eq.~(\ref{4cat}) with $k\equiv n \pmod{4}$. As all outcomes of the PNRD will herald a four-component cat state, the circuit is deterministic.  

\begin{figure}
    \centering
    \includegraphics{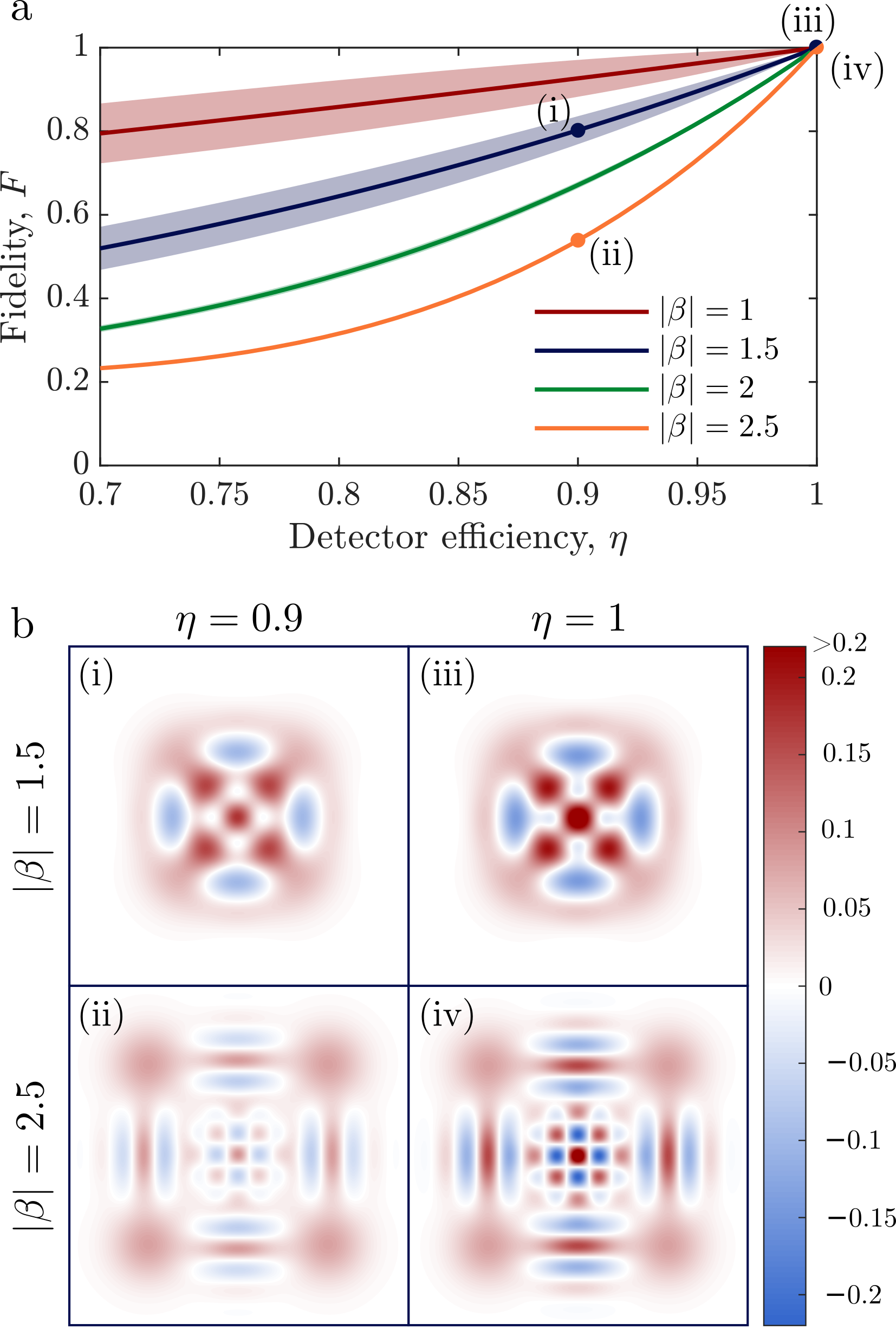}
    \caption{a) Fidelities between the actual and target four-component cat states for different input cat states as a function of the PNRD quantum efficiency. The fidelity is the mean fidelity for all measurement results, weighted according to the probability of obtaining each result, and the shaded areas show the standard deviation. b) Wigner functions for four different realizations as marked by (i--iv) in the upper figures.}
    \label{fig:DetectorEfficiency}
\end{figure}

\begin{figure}
    \centering
    \includegraphics{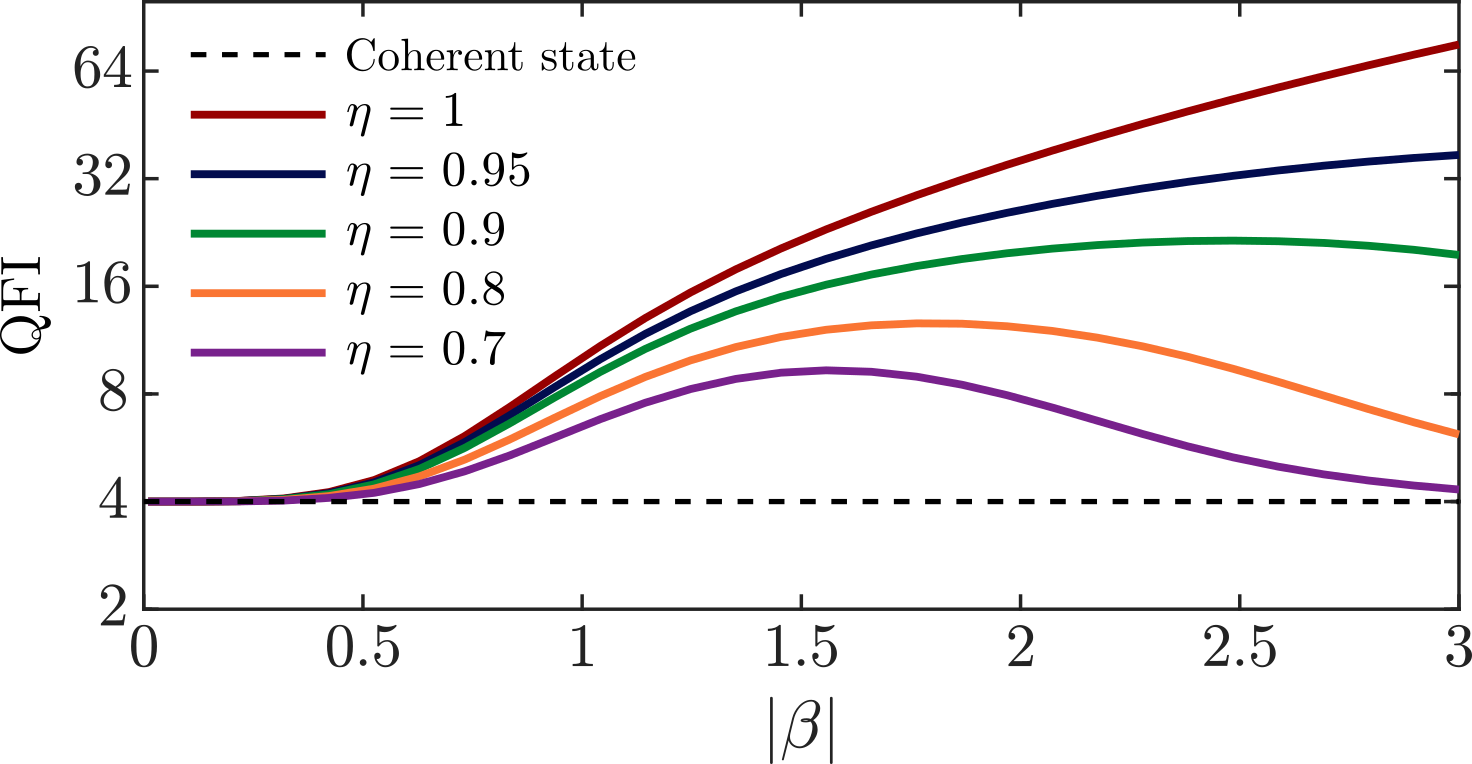}
    \caption{Quantum Fisher information with respect to phase space displacements defined as $\textrm{QFI}_\phi = 8\lim_{\varepsilon\rightarrow 0}\left(1-\sqrt{F(\rho_0,\rho_{\varepsilon,\phi})}\right)/\varepsilon^2$ with the displaced state $\rho_{\varepsilon,\phi}=\hat{D}(e^{i\phi}\varepsilon)\rho_0\hat{D}^\dagger(e^{i\phi}\varepsilon)$. For four-component cat states the QFI is independent of $\phi$.} 
    \label{QFI}
\end{figure}
We next examine the impact on the fidelity of a non-unity quantum efficiency of the PNRD. The PNRD with quantum efficiency $\eta$ is modelled by a perfect PNRD following a lossy channel with transmission $\eta$. For this detector we compute the fidelity, $F=\bra{\Phi_n}\rho\ket{\Phi_n}$, where $\rho$ is the generated state and $\ket{\Phi_n}$ is the target depending on the measurement result of the photon number $n\equiv0,1,2,3 \pmod{4}$. The resulting expected fidelities over all measurement outcomes (a numerical cut-off at $n=20$ was used for simulation) for four different input two-component cat states are shown in Fig.~\ref{fig:DetectorEfficiency}a. It is clear that a non-unity detector efficiency largely impacts the quality of the detected states, and it is therefore important to use a PNRD with very high efficiency. We note that there has been significant progress in developing high-efficiency PNRDs reaching nearly 100\% quantum efficiency~\cite{Lita2008,Humphreys2015}. Fig.~\ref{fig:DetectorEfficiency}b shows the Wigner functions of the output states when measuring $n\equiv 0 \textrm{ (mod 4)}$ photons with $\eta=0.9$ for $|\beta|=1.5$ (i) and $|\beta|=2.5$ (ii), as well as the corresponding pure states obtained when $\eta=1$ (iii, iv). For $|\beta|=1.5$, the phase-space features of the state are still clearly visible with imperfect detection. For $|\beta|=2.5$ the interference patterns are significantly dampened, particularly in the center of the state, but some negativity is still present. Thus, even though the fidelity is low (54\%), the characteristic phase-space features of the four-component cat state are still present.

Since larger photon number states are more sensitive to loss than smaller number photon states, one should expect the fidelity to depend on the measurement outcome. However, for $|\beta|\geq2$ the fidelity is practically independent of the measurement outcome, as seen in figure \ref{fig:DetectorEfficiency}a. This is because the increasing difficulty of detecting many photons with an imperfect detector, which would cause a lower fidelity for large $n$, is counteracted by the fact that the initial photon distribution decays exponentially for large $n$. 

For completeness, we also plot in Fig.~\ref{QFI} the state's quantum Fisher information (QFI) with respect to phase space displacements for different detector efficiencies. It represents the state's ability to sense phase space displacements \cite{zurek2001sub,duivenvoorden2017single} (irrespective of the direction) as the sensitivity scales as the QFI inverse. We note that for states generated with a non-unity efficiency PNRD, the sensitivity is optimized for a finite value of $|\beta|$. For comparison, the QFI of a coherent state is $4$, independent of the amplitude.

\begin{figure}
    \centering
    \includegraphics{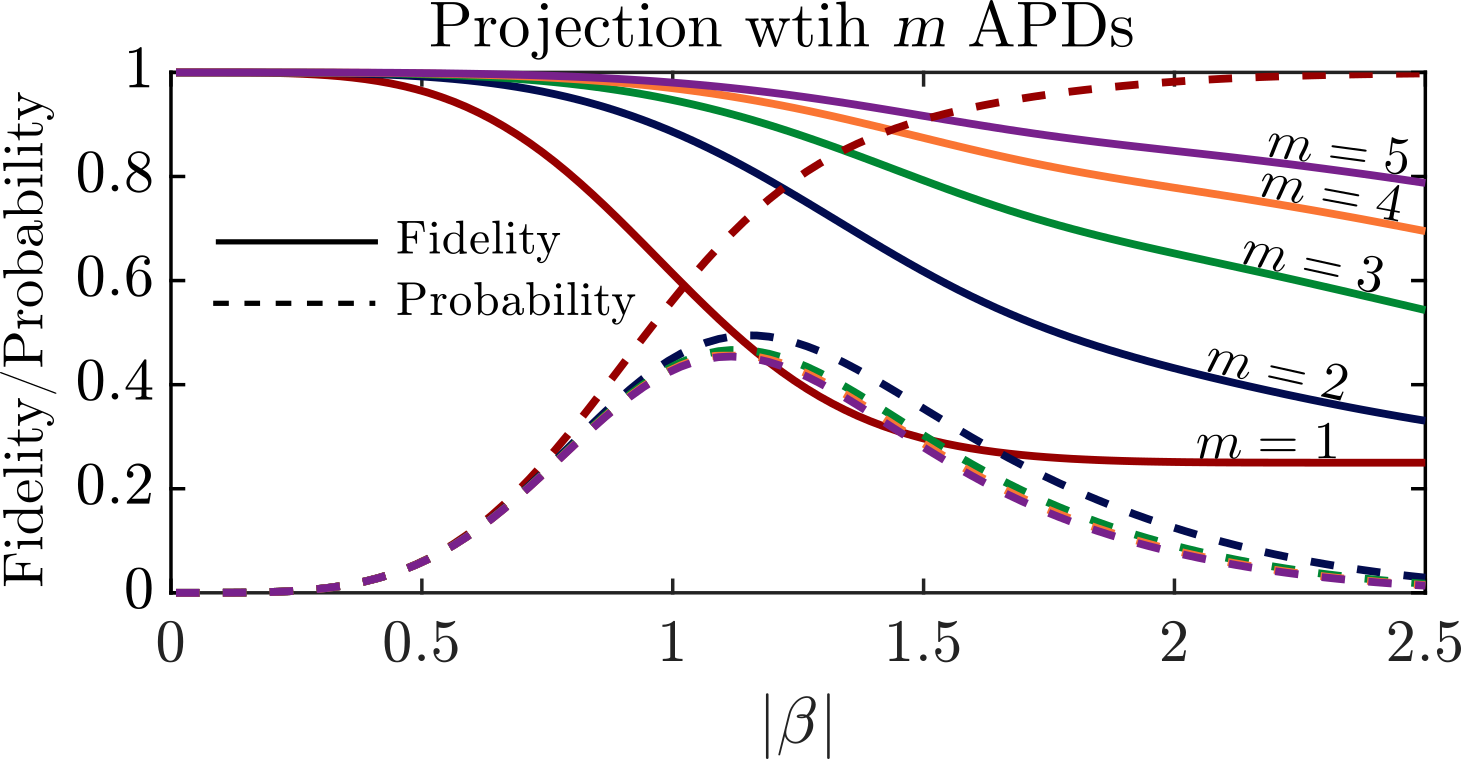}
    \caption{Fidelity and probability of the resulting state, post-selecting on a single 'on' event using $m$ on-off detectors (e.g. avalanche photodiodes (APDs)) as a function of the amplitude of the input cat states}
    \label{fig:APD}
\end{figure}
We have now seen that four-component cat states can be produced deterministically using a PNRD. Using an on-off photon detector, which is typically more experimentally feasible, one can still produce an exact four-component cat state by projecting onto the vacuum state, $\ket{n=0}_2$. The state is produced with a success rate of $P=e^{-|\beta|^2}\left(1+e^{-2|\beta|^2}+2e^{-|\beta|^2}\cos(|\beta|^2)\right)/(1+e^{-2|\beta|^2})^2$ employing an ideal on-off photon counter. One could also project onto vacuum using a heterodyne detector and post selecting on results near $(x,p)=(0,0)$. 

Since the probability of successfully projecting mode 2 onto the vacuum state decreases exponentially with $|\beta|$, it is also interesting to investigate the quality of the cat state when projecting onto the other outcome of the on-off photon detector as this will often be more probable. It is described by the projector $\Omega_{n>0}=I-|0\rangle\langle 0|$ and corresponds to a projection onto all Fock states except vacuum. Using such a measurement, the heralded output will contain a mixture of four different four-component cat states rendering the state mixed with the degree of mixedness determined by the amplitude of the input cat states. As a result, the fidelity drops rapidly as shown by the $m=1$ curve in Fig.~\ref{fig:APD}. For very low amplitudes, the output is fairly pure (and the fidelity high) while for amplitudes larger than $\sim 0.5$, the fidelity experiences a rapid decrease. This is explained by the increased occurrence of higher-order Fock states which the detector cannot discriminate. 

One can improve the fidelity by equally splitting mode 2 into $m$ modes and subsequently measuring each mode with an on-off photon detector \cite{achilles2004photon}. The POVM element corresponding to observing exactly 1 'on' click is $\sum_{n=1}^\infty m^{-(n-1)}\ket{n}\bra{n}$. As seen in figure \ref{fig:APD}, having multiple detectors allows for larger high fidelity cat states. The probability of observing exactly 1 'on' click is shown by the dashed lines, showing reasonable success probabilities for $|\beta|\in[0.5,2]$. Post-selecting on a higher number of clicks would similarly allow for even larger cat states with reasonable success probability given a sufficient number of detectors.
\begin{figure}
    \centering
    \includegraphics{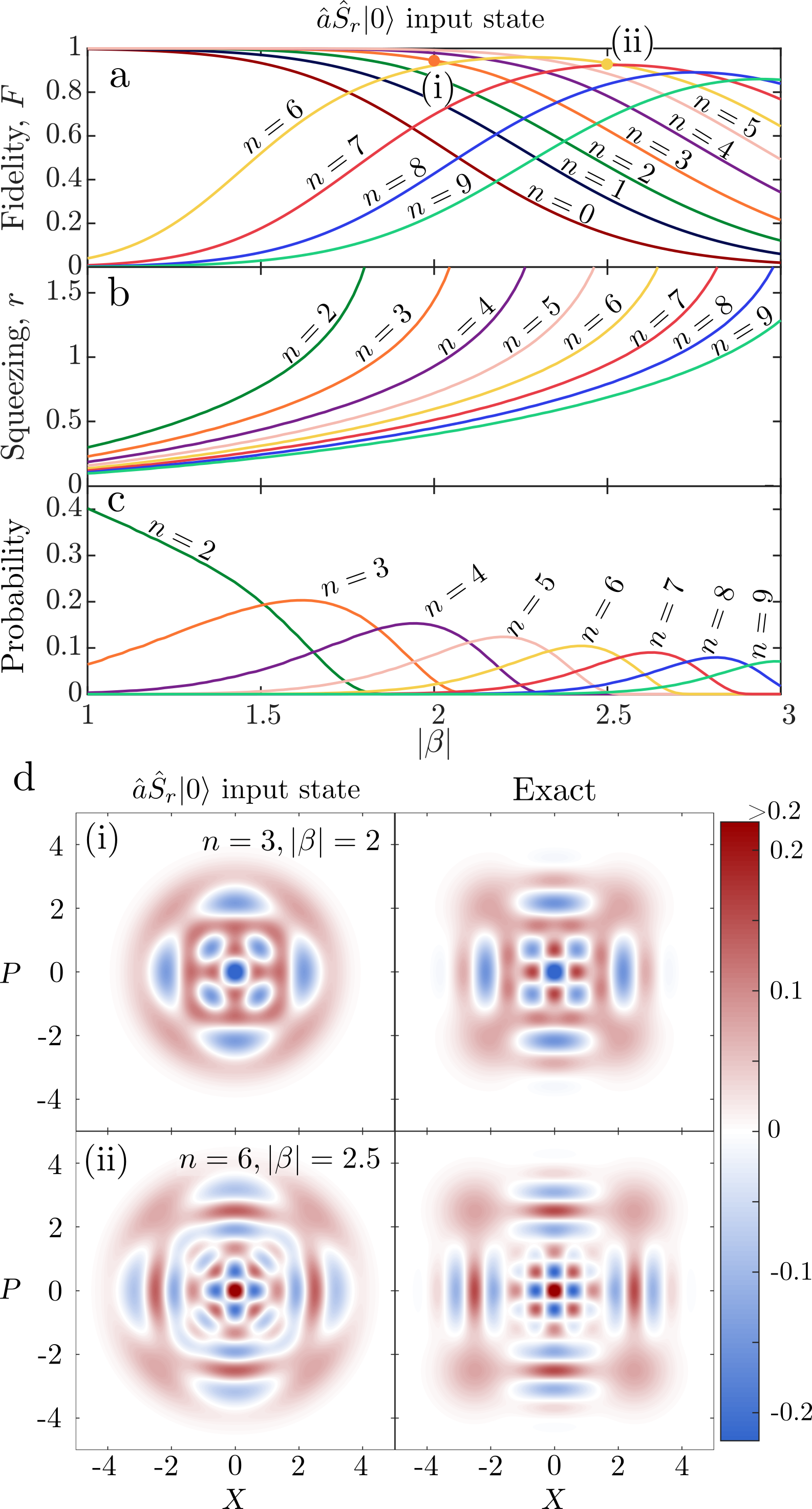}
    \caption{ a) Fidelity of output states relative to the ideal target state as a function of the amplitude of the target state where the input states to the circuit are photon-subtracted squeezed states. b) The squeezing parameters $r$ of the input states are chosen for each $\beta$ to optimize the fidelity. c) The probability of measuring $n$ photons for the corresponding optimum squeezing parameter, $r$. d) Left: Wigner functions of the output states marked by (i) and (ii) in a) and b). Right: corresponding Wigner functions of the exact 4-component cat state target.}
    \label{fig:kitten}
\end{figure}

We have now shown that exact four-component cat states can be produced using a simple circuit if we have at our disposal a pair of two-component cat states. The two-component cat states can be produced deterministically using the strong interaction between a single atom and light as recently demonstrated~\cite{Hacker2019}. However, knowing that single photon-subtracted squeezed states resemble two-component cat states \cite{dakna1997generating}, it is interesting to investigate the possibility of using such states as inputs to the circuit for the generation of approximate four-component cat states. Using two single photon-subtracted squeezed vacuum states, $\hat{a}_1\hat{S}_1\ket{0}$ and $\hat{a}_2\hat{S}_2
^\dagger\ket{0}$ (where $\hat{a}_i$ is the annihilation operator for mode $i=1,2$ and $\hat{S}_i=e^{r/2(\hat{a}_i^2-\hat{a}_i^{\dagger 2})}$ is the squeezing operator with $r$ being the squeezing parameter), as the input cat states, the state after the beam splitter reads
\begin{multline}
    \hat{U}_\textrm{BS}\left(\hat{a}_1\hat{S}_1\hat{a}_2\hat{S}_2^\dagger\ket{0}_1\ket{0}_2\right) \\
    = \frac{1}{2}\frac{1}{\cosh(r)}\sum_n \tanh^n(r)\bigg(\sqrt{n(n-1)}\ket{n-2}_1  \\
    - \sqrt{(n+2)(n+1)}\tanh^2(r)\ket{n+2}_1\bigg)\ket{n}_2
    \label{eq:kitten}
\end{multline}
where we have used the equality $\hat{U}_\textrm{BS}\hat{a}_1\hat{a}_2 = \frac{1}{2}(\hat{a}_1^2 - \hat{a}_2^2)\hat{U}_\textrm{BS}$. It is clear that the scheme will herald a two-photon Fock state, $\ket{2}_1$, when projecting on $\ket{0}_2$ and a three-photon Fock state, $\ket{3}_1$, when projecting on $\ket{1}_2$. It is, however, more interesting to project onto even higher Fock states as this will herald Fock state superpositions, e.g. the (unnormalized) states $\sqrt{2}\ket{0}_1-\sqrt{6}\tanh^2(r) \ket{4}_1$ and $\sqrt{6}\ket{1}_1-\sqrt{20}\tanh^2(r)\ket{5}_1$ are produced when the PNRD counts 2 and 3 photons, respectively. In Fig.~\ref{fig:kitten}a we present the fidelity of these states with respect to the ideal four component cat states for different photon counting measurement outcomes from 0 to 9 photons. In these plots we have optimized the squeezing parameter for each realization to maximize the fidelity, with the optimized values shown in Fig.~\ref{fig:kitten}b and corresponding probability of measuring $n$ photons shown in Fig. \ref{fig:kitten}c. Note, that all measurement results of more than 1 photon yield a state which approximates a four-component cat state to some degree, according to \eqref{eq:kitten}, even if squeezing parameter is chosen to optimize the fidelity for a specific outcome. In experiment, one might therefore post-select on several measurement outcomes. The abrupt drop in fidelity at small $|\beta|$ for $n\geq 6$ is due to the lowest Fock term missing from the output state, compared to the exact four-component cat state, as seen in \eqref{eq:kitten}. Fig.~\ref{fig:kitten}d shows the Wigner function of generated approximate states marked by (i) and (ii) in Fig.~\ref{fig:kitten}a, in comparison to the exact target states.

As shown in \cite{dakna1997generating}, better two-component cat states can be produced by subtracting multiple photons from squeezed states. From numerical analysis we have found that using such states as input will also result in even higher fidelity output states. However, it is an open question whether arbitrarily large high-fidelity 4-component cat states can be produced with this approach.

In conclusion, we have proposed a simple circuit for the generation of four-component cat states which eventually could be used for fault-tolerant quantum computing and communication. The scheme is deterministic and exact if two-component cat states and photon-number-resolving detectors are available.
\\ \\
\noindent\textbf{Funding.} Danish National Research Foundation through the Center of Excellence for Macroscopic Quantum States (bigQ, DNRF142).
\\ \\
\noindent\textbf{Disclosures.} The authors declare no conflicts of interest.

\bibliography{literature}

%Control: production of eprint (-1) disabled
\begin{thebibliography}{35}%
\makeatletter
\providecommand \@ifxundefined [1]{%
 \@ifx{#1\undefined}
}%
\providecommand \@ifnum [1]{%
 \ifnum #1\expandafter \@firstoftwo
 \else \expandafter \@secondoftwo
 \fi
}%
\providecommand \@ifx [1]{%
 \ifx #1\expandafter \@firstoftwo
 \else \expandafter \@secondoftwo
 \fi
}%
\providecommand \natexlab [1]{#1}%
\providecommand \enquote  [1]{``#1''}%
\providecommand \bibnamefont  [1]{#1}%
\providecommand \bibfnamefont [1]{#1}%
\providecommand \citenamefont [1]{#1}%
\providecommand \href@noop [0]{\@secondoftwo}%
\providecommand \href [0]{\begingroup \@sanitize@url \@href}%
\providecommand \@href[1]{\@@startlink{#1}\@@href}%
\providecommand \@@href[1]{\endgroup#1\@@endlink}%
\providecommand \@sanitize@url [0]{\catcode `\\12\catcode `\$12\catcode
  `\&12\catcode `\#12\catcode `\^12\catcode `\_12\catcode `\%12\relax}%
\providecommand \@@startlink[1]{}%
\providecommand \@@endlink[0]{}%
\providecommand \url  [0]{\begingroup\@sanitize@url \@url }%
\providecommand \@url [1]{\endgroup\@href {#1}{\urlprefix }}%
\providecommand \urlprefix  [0]{URL }%
\providecommand \Eprint [0]{\href }%
\providecommand \doibase [0]{http://dx.doi.org/}%
\providecommand \selectlanguage [0]{\@gobble}%
\providecommand \bibinfo  [0]{\@secondoftwo}%
\providecommand \bibfield  [0]{\@secondoftwo}%
\providecommand \translation [1]{[#1]}%
\providecommand \BibitemOpen [0]{}%
\providecommand \bibitemStop [0]{}%
\providecommand \bibitemNoStop [0]{.\EOS\space}%
\providecommand \EOS [0]{\spacefactor3000\relax}%
\providecommand \BibitemShut  [1]{\csname bibitem#1\endcsname}%
\let\auto@bib@innerbib\@empty
%</preamble>
\bibitem [{\citenamefont {Gottesman}\ \emph {et~al.}(2001)\citenamefont
  {Gottesman}, \citenamefont {Kitaev},\ and\ \citenamefont
  {Preskill}}]{Gottesman2001}%
  \BibitemOpen
  \bibfield  {author} {\bibinfo {author} {\bibfnamefont {D.}~\bibnamefont
  {Gottesman}}, \bibinfo {author} {\bibfnamefont {A.}~\bibnamefont {Kitaev}}, \
  and\ \bibinfo {author} {\bibfnamefont {J.}~\bibnamefont {Preskill}},\ }\href
  {\doibase 10.1103/PhysRevA.64.012310} {\bibfield  {journal} {\bibinfo
  {journal} {Physical Review A}\ }\textbf {\bibinfo {volume} {64}},\ \bibinfo
  {pages} {012310} (\bibinfo {year} {2001})}\BibitemShut {NoStop}%
\bibitem [{\citenamefont {Cochrane}\ \emph {et~al.}(1999)\citenamefont
  {Cochrane}, \citenamefont {Milburn},\ and\ \citenamefont
  {Munro}}]{Cochrane1999}%
  \BibitemOpen
  \bibfield  {author} {\bibinfo {author} {\bibfnamefont {P.~T.}\ \bibnamefont
  {Cochrane}}, \bibinfo {author} {\bibfnamefont {G.~J.}\ \bibnamefont
  {Milburn}}, \ and\ \bibinfo {author} {\bibfnamefont {W.~J.}\ \bibnamefont
  {Munro}},\ }\href@noop {} {\bibfield  {journal} {\bibinfo  {journal}
  {Physical Review A}\ }\textbf {\bibinfo {volume} {59}},\ \bibinfo {pages}
  {2631} (\bibinfo {year} {1999})}\BibitemShut {NoStop}%
\bibitem [{\citenamefont {Leghtas}\ \emph {et~al.}(2013)\citenamefont
  {Leghtas}, \citenamefont {Kirchmair}, \citenamefont {Vlastakis},
  \citenamefont {Schoelkopf}, \citenamefont {Devoret},\ and\ \citenamefont
  {Mirrahimi}}]{leghtas2013hardware}%
  \BibitemOpen
  \bibfield  {author} {\bibinfo {author} {\bibfnamefont {Z.}~\bibnamefont
  {Leghtas}}, \bibinfo {author} {\bibfnamefont {G.}~\bibnamefont {Kirchmair}},
  \bibinfo {author} {\bibfnamefont {B.}~\bibnamefont {Vlastakis}}, \bibinfo
  {author} {\bibfnamefont {R.~J.}\ \bibnamefont {Schoelkopf}}, \bibinfo
  {author} {\bibfnamefont {M.~H.}\ \bibnamefont {Devoret}}, \ and\ \bibinfo
  {author} {\bibfnamefont {M.}~\bibnamefont {Mirrahimi}},\ }\href@noop {}
  {\bibfield  {journal} {\bibinfo  {journal} {Physical Review Letters}\
  }\textbf {\bibinfo {volume} {111}},\ \bibinfo {pages} {120501} (\bibinfo
  {year} {2013})}\BibitemShut {NoStop}%
\bibitem [{\citenamefont {Li}\ \emph {et~al.}(2017)\citenamefont {Li},
  \citenamefont {Zou}, \citenamefont {Albert}, \citenamefont {Muralidharan},
  \citenamefont {Girvin},\ and\ \citenamefont {Jiang}}]{Li2017}%
  \BibitemOpen
  \bibfield  {author} {\bibinfo {author} {\bibfnamefont {L.}~\bibnamefont
  {Li}}, \bibinfo {author} {\bibfnamefont {C.~L.}\ \bibnamefont {Zou}},
  \bibinfo {author} {\bibfnamefont {V.~V.}\ \bibnamefont {Albert}}, \bibinfo
  {author} {\bibfnamefont {S.}~\bibnamefont {Muralidharan}}, \bibinfo {author}
  {\bibfnamefont {S.~M.}\ \bibnamefont {Girvin}}, \ and\ \bibinfo {author}
  {\bibfnamefont {L.}~\bibnamefont {Jiang}},\ }\href {\doibase
  10.1103/PhysRevLett.119.030502} {\bibfield  {journal} {\bibinfo  {journal}
  {Physical Review Letters}\ }\textbf {\bibinfo {volume} {119}},\ \bibinfo
  {pages} {030502} (\bibinfo {year} {2017})}\BibitemShut {NoStop}%
\bibitem [{\citenamefont {Michael}\ \emph {et~al.}(2016)\citenamefont
  {Michael}, \citenamefont {Silveri}, \citenamefont {Brierley}, \citenamefont
  {Albert}, \citenamefont {Salmilehto}, \citenamefont {Jiang},\ and\
  \citenamefont {Girvin}}]{Michael2016}%
  \BibitemOpen
  \bibfield  {author} {\bibinfo {author} {\bibfnamefont {M.~H.}\ \bibnamefont
  {Michael}}, \bibinfo {author} {\bibfnamefont {M.}~\bibnamefont {Silveri}},
  \bibinfo {author} {\bibfnamefont {R.~T.}\ \bibnamefont {Brierley}}, \bibinfo
  {author} {\bibfnamefont {V.~V.}\ \bibnamefont {Albert}}, \bibinfo {author}
  {\bibfnamefont {J.}~\bibnamefont {Salmilehto}}, \bibinfo {author}
  {\bibfnamefont {L.}~\bibnamefont {Jiang}}, \ and\ \bibinfo {author}
  {\bibfnamefont {S.~M.}\ \bibnamefont {Girvin}},\ }\href {\doibase
  10.1103/PhysRevX.6.031006} {\bibfield  {journal} {\bibinfo  {journal}
  {Physical Review X}\ }\textbf {\bibinfo {volume} {6}},\ \bibinfo {pages} {1}
  (\bibinfo {year} {2016})}\BibitemShut {NoStop}%
\bibitem [{\citenamefont {Albert}\ \emph {et~al.}(2018)\citenamefont {Albert},
  \citenamefont {Noh}, \citenamefont {Duivenvoorden}, \citenamefont {Young},
  \citenamefont {Brierley}, \citenamefont {Reinhold}, \citenamefont {Vuillot},
  \citenamefont {Li}, \citenamefont {Shen}, \citenamefont {Girvin},
  \citenamefont {Terhal},\ and\ \citenamefont {Jiang}}]{Albert2018}%
  \BibitemOpen
  \bibfield  {author} {\bibinfo {author} {\bibfnamefont {V.~V.}\ \bibnamefont
  {Albert}}, \bibinfo {author} {\bibfnamefont {K.}~\bibnamefont {Noh}},
  \bibinfo {author} {\bibfnamefont {K.}~\bibnamefont {Duivenvoorden}}, \bibinfo
  {author} {\bibfnamefont {D.~J.}\ \bibnamefont {Young}}, \bibinfo {author}
  {\bibfnamefont {R.~T.}\ \bibnamefont {Brierley}}, \bibinfo {author}
  {\bibfnamefont {P.}~\bibnamefont {Reinhold}}, \bibinfo {author}
  {\bibfnamefont {C.}~\bibnamefont {Vuillot}}, \bibinfo {author} {\bibfnamefont
  {L.}~\bibnamefont {Li}}, \bibinfo {author} {\bibfnamefont {C.}~\bibnamefont
  {Shen}}, \bibinfo {author} {\bibfnamefont {S.~M.}\ \bibnamefont {Girvin}},
  \bibinfo {author} {\bibfnamefont {B.~M.}\ \bibnamefont {Terhal}}, \ and\
  \bibinfo {author} {\bibfnamefont {L.}~\bibnamefont {Jiang}},\ }\href
  {\doibase 10.1103/PhysRevA.97.032346} {\bibfield  {journal} {\bibinfo
  {journal} {Physical Review A}\ }\textbf {\bibinfo {volume} {97}},\ \bibinfo
  {pages} {032346} (\bibinfo {year} {2018})}\BibitemShut {NoStop}%
\bibitem [{\citenamefont {Grimsmo}\ \emph {et~al.}(2019)\citenamefont
  {Grimsmo}, \citenamefont {Combes},\ and\ \citenamefont
  {Baragiola}}]{grimsmo2019quantum}%
  \BibitemOpen
  \bibfield  {author} {\bibinfo {author} {\bibfnamefont {A.~L.}\ \bibnamefont
  {Grimsmo}}, \bibinfo {author} {\bibfnamefont {J.}~\bibnamefont {Combes}}, \
  and\ \bibinfo {author} {\bibfnamefont {B.~Q.}\ \bibnamefont {Baragiola}},\
  }\href {https://arxiv.org/abs/1901.08071} {\bibfield  {journal} {\bibinfo
  {journal} {arXiv preprint arXiv:1901.08071}\ } (\bibinfo {year}
  {2019})}\BibitemShut {NoStop}%
\bibitem [{\citenamefont {Ofek}\ \emph {et~al.}(2016)\citenamefont {Ofek},
  \citenamefont {Petrenko}, \citenamefont {Heeres}, \citenamefont {Reinhold},
  \citenamefont {Leghtas}, \citenamefont {Vlastakis}, \citenamefont {Liu},
  \citenamefont {Frunzio}, \citenamefont {Girvin}, \citenamefont {Jiang},
  \citenamefont {Mirrahimi}, \citenamefont {Devoret},\ and\ \citenamefont
  {Schoelkopf}}]{Ofek2016a}%
  \BibitemOpen
  \bibfield  {author} {\bibinfo {author} {\bibfnamefont {N.}~\bibnamefont
  {Ofek}}, \bibinfo {author} {\bibfnamefont {A.}~\bibnamefont {Petrenko}},
  \bibinfo {author} {\bibfnamefont {R.}~\bibnamefont {Heeres}}, \bibinfo
  {author} {\bibfnamefont {P.}~\bibnamefont {Reinhold}}, \bibinfo {author}
  {\bibfnamefont {Z.}~\bibnamefont {Leghtas}}, \bibinfo {author} {\bibfnamefont
  {B.}~\bibnamefont {Vlastakis}}, \bibinfo {author} {\bibfnamefont
  {Y.}~\bibnamefont {Liu}}, \bibinfo {author} {\bibfnamefont {L.}~\bibnamefont
  {Frunzio}}, \bibinfo {author} {\bibfnamefont {S.~M.}\ \bibnamefont {Girvin}},
  \bibinfo {author} {\bibfnamefont {L.}~\bibnamefont {Jiang}}, \bibinfo
  {author} {\bibfnamefont {M.}~\bibnamefont {Mirrahimi}}, \bibinfo {author}
  {\bibfnamefont {M.~H.}\ \bibnamefont {Devoret}}, \ and\ \bibinfo {author}
  {\bibfnamefont {R.~J.}\ \bibnamefont {Schoelkopf}},\ }\href {\doibase
  10.1038/nature18949} {\bibfield  {journal} {\bibinfo  {journal} {Nature}\
  }\textbf {\bibinfo {volume} {536}},\ \bibinfo {pages} {441} (\bibinfo {year}
  {2016})}\BibitemShut {NoStop}%
\bibitem [{\citenamefont {Hu}\ \emph {et~al.}(2019)\citenamefont {Hu},
  \citenamefont {Ma}, \citenamefont {Cai}, \citenamefont {Mu}, \citenamefont
  {Xu}, \citenamefont {Wang}, \citenamefont {Wu}, \citenamefont {Wang},
  \citenamefont {Song}, \citenamefont {Zou}, \citenamefont {Girvin},
  \citenamefont {Duan},\ and\ \citenamefont {Sun}}]{hu2019}%
  \BibitemOpen
  \bibfield  {author} {\bibinfo {author} {\bibfnamefont {L.}~\bibnamefont
  {Hu}}, \bibinfo {author} {\bibfnamefont {Y.}~\bibnamefont {Ma}}, \bibinfo
  {author} {\bibfnamefont {W.}~\bibnamefont {Cai}}, \bibinfo {author}
  {\bibfnamefont {X.}~\bibnamefont {Mu}}, \bibinfo {author} {\bibfnamefont
  {Y.}~\bibnamefont {Xu}}, \bibinfo {author} {\bibfnamefont {W.}~\bibnamefont
  {Wang}}, \bibinfo {author} {\bibfnamefont {Y.}~\bibnamefont {Wu}}, \bibinfo
  {author} {\bibfnamefont {H.}~\bibnamefont {Wang}}, \bibinfo {author}
  {\bibfnamefont {Y.~P.}\ \bibnamefont {Song}}, \bibinfo {author}
  {\bibfnamefont {C.~L.}\ \bibnamefont {Zou}}, \bibinfo {author} {\bibfnamefont
  {S.~M.}\ \bibnamefont {Girvin}}, \bibinfo {author} {\bibfnamefont {L.~M.}\
  \bibnamefont {Duan}}, \ and\ \bibinfo {author} {\bibfnamefont
  {L.}~\bibnamefont {Sun}},\ }\href {\doibase 10.1038/s41567-018-0414-3}
  {\bibfield  {journal} {\bibinfo  {journal} {Nature Physics}\ }\textbf
  {\bibinfo {volume} {15}},\ \bibinfo {pages} {503} (\bibinfo {year}
  {2019})}\BibitemShut {NoStop}%
\bibitem [{\citenamefont {Campagne-Ibarcq}\ \emph {et~al.}(2019)\citenamefont
  {Campagne-Ibarcq}, \citenamefont {Eickbusch}, \citenamefont {Touzard},
  \citenamefont {Zalys-Geller}, \citenamefont {Frattini}, \citenamefont
  {Sivak}, \citenamefont {Reinhold}, \citenamefont {Puri}, \citenamefont
  {Shankar}, \citenamefont {Schoelkopf}, \citenamefont {Frunzio}, \citenamefont
  {Mirrahimi},\ and\ \citenamefont {Devoret}}]{Campagne-Ibarcq2019}%
  \BibitemOpen
  \bibfield  {author} {\bibinfo {author} {\bibfnamefont {P.}~\bibnamefont
  {Campagne-Ibarcq}}, \bibinfo {author} {\bibfnamefont {A.}~\bibnamefont
  {Eickbusch}}, \bibinfo {author} {\bibfnamefont {S.}~\bibnamefont {Touzard}},
  \bibinfo {author} {\bibfnamefont {E.}~\bibnamefont {Zalys-Geller}}, \bibinfo
  {author} {\bibfnamefont {N.~E.}\ \bibnamefont {Frattini}}, \bibinfo {author}
  {\bibfnamefont {V.~V.}\ \bibnamefont {Sivak}}, \bibinfo {author}
  {\bibfnamefont {P.}~\bibnamefont {Reinhold}}, \bibinfo {author}
  {\bibfnamefont {S.}~\bibnamefont {Puri}}, \bibinfo {author} {\bibfnamefont
  {S.}~\bibnamefont {Shankar}}, \bibinfo {author} {\bibfnamefont {R.~J.}\
  \bibnamefont {Schoelkopf}}, \bibinfo {author} {\bibfnamefont
  {L.}~\bibnamefont {Frunzio}}, \bibinfo {author} {\bibfnamefont
  {M.}~\bibnamefont {Mirrahimi}}, \ and\ \bibinfo {author} {\bibfnamefont
  {M.~H.}\ \bibnamefont {Devoret}},\ }\href {http://arxiv.org/abs/1907.12487}
  {\bibfield  {journal} {\bibinfo  {journal} {arXiv preprint arXiv:1907.12487}\
  } (\bibinfo {year} {2019})}\BibitemShut {NoStop}%
\bibitem [{\citenamefont {Fl{\"{u}}hmann}\ \emph {et~al.}(2018)\citenamefont
  {Fl{\"{u}}hmann}, \citenamefont {Negnevitsky}, \citenamefont {Marinelli},\
  and\ \citenamefont {Home}}]{Fluhmann2018}%
  \BibitemOpen
  \bibfield  {author} {\bibinfo {author} {\bibfnamefont {C.}~\bibnamefont
  {Fl{\"{u}}hmann}}, \bibinfo {author} {\bibfnamefont {V.}~\bibnamefont
  {Negnevitsky}}, \bibinfo {author} {\bibfnamefont {M.}~\bibnamefont
  {Marinelli}}, \ and\ \bibinfo {author} {\bibfnamefont {J.~P.}\ \bibnamefont
  {Home}},\ }\href {\doibase 10.1103/PhysRevX.8.021001} {\bibfield  {journal}
  {\bibinfo  {journal} {Physical Review X}\ }\textbf {\bibinfo {volume} {8}},\
  \bibinfo {pages} {021001} (\bibinfo {year} {2018})}\BibitemShut {NoStop}%
\bibitem [{\citenamefont {Fl{\"{u}}hmann}\ \emph {et~al.}(2019)\citenamefont
  {Fl{\"{u}}hmann}, \citenamefont {Nguyen}, \citenamefont {Marinelli},
  \citenamefont {Negnevitsky}, \citenamefont {Mehta},\ and\ \citenamefont
  {Home}}]{Fluhmann2019}%
  \BibitemOpen
  \bibfield  {author} {\bibinfo {author} {\bibfnamefont {C.}~\bibnamefont
  {Fl{\"{u}}hmann}}, \bibinfo {author} {\bibfnamefont {T.~L.}\ \bibnamefont
  {Nguyen}}, \bibinfo {author} {\bibfnamefont {M.}~\bibnamefont {Marinelli}},
  \bibinfo {author} {\bibfnamefont {V.}~\bibnamefont {Negnevitsky}}, \bibinfo
  {author} {\bibfnamefont {K.}~\bibnamefont {Mehta}}, \ and\ \bibinfo {author}
  {\bibfnamefont {J.~P.}\ \bibnamefont {Home}},\ }\href {\doibase
  10.1038/s41586-019-0960-6} {\bibfield  {journal} {\bibinfo  {journal}
  {Nature}\ }\textbf {\bibinfo {volume} {566}},\ \bibinfo {pages} {513}
  (\bibinfo {year} {2019})}\BibitemShut {NoStop}%
\bibitem [{\citenamefont {Weedbrook}\ \emph {et~al.}(2012)\citenamefont
  {Weedbrook}, \citenamefont {Pirandola}, \citenamefont
  {Garc{\'{i}}a-Patr{\'{o}}n}, \citenamefont {Cerf}, \citenamefont {Ralph},
  \citenamefont {Shapiro},\ and\ \citenamefont {Lloyd}}]{Weedbrook2012}%
  \BibitemOpen
  \bibfield  {author} {\bibinfo {author} {\bibfnamefont {C.}~\bibnamefont
  {Weedbrook}}, \bibinfo {author} {\bibfnamefont {S.}~\bibnamefont
  {Pirandola}}, \bibinfo {author} {\bibfnamefont {R.}~\bibnamefont
  {Garc{\'{i}}a-Patr{\'{o}}n}}, \bibinfo {author} {\bibfnamefont {N.~J.}\
  \bibnamefont {Cerf}}, \bibinfo {author} {\bibfnamefont {T.~C.}\ \bibnamefont
  {Ralph}}, \bibinfo {author} {\bibfnamefont {J.~H.}\ \bibnamefont {Shapiro}},
  \ and\ \bibinfo {author} {\bibfnamefont {S.}~\bibnamefont {Lloyd}},\ }\href
  {\doibase 10.1103/RevModPhys.84.621} {\bibfield  {journal} {\bibinfo
  {journal} {Reviews of Modern Physics}\ }\textbf {\bibinfo {volume} {84}},\
  \bibinfo {pages} {621} (\bibinfo {year} {2012})}\BibitemShut {NoStop}%
\bibitem [{\citenamefont {Muralidharan}\ \emph {et~al.}(2016)\citenamefont
  {Muralidharan}, \citenamefont {Li}, \citenamefont {Kim}, \citenamefont
  {L{\"u}tkenhaus}, \citenamefont {Lukin},\ and\ \citenamefont
  {Jiang}}]{muralidharan2016optimal}%
  \BibitemOpen
  \bibfield  {author} {\bibinfo {author} {\bibfnamefont {S.}~\bibnamefont
  {Muralidharan}}, \bibinfo {author} {\bibfnamefont {L.}~\bibnamefont {Li}},
  \bibinfo {author} {\bibfnamefont {J.}~\bibnamefont {Kim}}, \bibinfo {author}
  {\bibfnamefont {N.}~\bibnamefont {L{\"u}tkenhaus}}, \bibinfo {author}
  {\bibfnamefont {M.~D.}\ \bibnamefont {Lukin}}, \ and\ \bibinfo {author}
  {\bibfnamefont {L.}~\bibnamefont {Jiang}},\ }\href@noop {} {\bibfield
  {journal} {\bibinfo  {journal} {Scientific Reports}\ }\textbf {\bibinfo
  {volume} {6}},\ \bibinfo {pages} {20463} (\bibinfo {year}
  {2016})}\BibitemShut {NoStop}%
\bibitem [{\citenamefont {Pysher}\ \emph {et~al.}(2011)\citenamefont {Pysher},
  \citenamefont {Miwa}, \citenamefont {Shahrokhshahi}, \citenamefont
  {Bloomer},\ and\ \citenamefont {Pfister}}]{pysher2011parallel}%
  \BibitemOpen
  \bibfield  {author} {\bibinfo {author} {\bibfnamefont {M.}~\bibnamefont
  {Pysher}}, \bibinfo {author} {\bibfnamefont {Y.}~\bibnamefont {Miwa}},
  \bibinfo {author} {\bibfnamefont {R.}~\bibnamefont {Shahrokhshahi}}, \bibinfo
  {author} {\bibfnamefont {R.}~\bibnamefont {Bloomer}}, \ and\ \bibinfo
  {author} {\bibfnamefont {O.}~\bibnamefont {Pfister}},\ }\href
  {https://journals.aps.org/prl/abstract/10.1103/PhysRevLett.107.030505}
  {\bibfield  {journal} {\bibinfo  {journal} {Physical review letters}\
  }\textbf {\bibinfo {volume} {107}},\ \bibinfo {pages} {030505} (\bibinfo
  {year} {2011})}\BibitemShut {NoStop}%
\bibitem [{\citenamefont {Yokoyama}\ \emph {et~al.}(2013)\citenamefont
  {Yokoyama}, \citenamefont {Ukai}, \citenamefont {Armstrong}, \citenamefont
  {Sornphiphatphong}, \citenamefont {Kaji}, \citenamefont {Suzuki},
  \citenamefont {Yoshikawa}, \citenamefont {Yonezawa}, \citenamefont
  {Menicucci},\ and\ \citenamefont {Furusawa}}]{Yokoyama2013}%
  \BibitemOpen
  \bibfield  {author} {\bibinfo {author} {\bibfnamefont {S.}~\bibnamefont
  {Yokoyama}}, \bibinfo {author} {\bibfnamefont {R.}~\bibnamefont {Ukai}},
  \bibinfo {author} {\bibfnamefont {S.~C.}\ \bibnamefont {Armstrong}}, \bibinfo
  {author} {\bibfnamefont {C.}~\bibnamefont {Sornphiphatphong}}, \bibinfo
  {author} {\bibfnamefont {T.}~\bibnamefont {Kaji}}, \bibinfo {author}
  {\bibfnamefont {S.}~\bibnamefont {Suzuki}}, \bibinfo {author} {\bibfnamefont
  {J.~I.}\ \bibnamefont {Yoshikawa}}, \bibinfo {author} {\bibfnamefont
  {H.}~\bibnamefont {Yonezawa}}, \bibinfo {author} {\bibfnamefont {N.~C.}\
  \bibnamefont {Menicucci}}, \ and\ \bibinfo {author} {\bibfnamefont
  {A.}~\bibnamefont {Furusawa}},\ }\href {\doibase 10.1038/nphoton.2013.287}
  {\bibfield  {journal} {\bibinfo  {journal} {Nature Photonics}\ }\textbf
  {\bibinfo {volume} {7}},\ \bibinfo {pages} {982} (\bibinfo {year}
  {2013})}\BibitemShut {NoStop}%
\bibitem [{\citenamefont {Asavanant}\ \emph {et~al.}(2019)\citenamefont
  {Asavanant}, \citenamefont {Shiozawa}, \citenamefont {Yokoyama},
  \citenamefont {Charoensombutamon}, \citenamefont {Emura}, \citenamefont
  {Alexander}, \citenamefont {Takeda}, \citenamefont {Yoshikawa}, \citenamefont
  {Menicucci}, \citenamefont {Yonezawa},\ and\ \citenamefont
  {Furusawa}}]{Asavanant2019}%
  \BibitemOpen
  \bibfield  {author} {\bibinfo {author} {\bibfnamefont {W.}~\bibnamefont
  {Asavanant}}, \bibinfo {author} {\bibfnamefont {Y.}~\bibnamefont {Shiozawa}},
  \bibinfo {author} {\bibfnamefont {S.}~\bibnamefont {Yokoyama}}, \bibinfo
  {author} {\bibfnamefont {B.}~\bibnamefont {Charoensombutamon}}, \bibinfo
  {author} {\bibfnamefont {H.}~\bibnamefont {Emura}}, \bibinfo {author}
  {\bibfnamefont {R.~N.}\ \bibnamefont {Alexander}}, \bibinfo {author}
  {\bibfnamefont {S.}~\bibnamefont {Takeda}}, \bibinfo {author} {\bibfnamefont
  {J.-i.}\ \bibnamefont {Yoshikawa}}, \bibinfo {author} {\bibfnamefont {N.~C.}\
  \bibnamefont {Menicucci}}, \bibinfo {author} {\bibfnamefont {H.}~\bibnamefont
  {Yonezawa}}, \ and\ \bibinfo {author} {\bibfnamefont {A.}~\bibnamefont
  {Furusawa}},\ }\href {\doibase 10.1126/science.aay2645} {\bibfield  {journal}
  {\bibinfo  {journal} {Science}\ }\textbf {\bibinfo {volume} {366}},\ \bibinfo
  {pages} {373} (\bibinfo {year} {2019})}\BibitemShut {NoStop}%
\bibitem [{\citenamefont {Larsen}\ \emph {et~al.}(2019)\citenamefont {Larsen},
  \citenamefont {Guo}, \citenamefont {Breum}, \citenamefont
  {Neergaard-Nielsen},\ and\ \citenamefont {Andersen}}]{Larsen2019}%
  \BibitemOpen
  \bibfield  {author} {\bibinfo {author} {\bibfnamefont {M.~V.}\ \bibnamefont
  {Larsen}}, \bibinfo {author} {\bibfnamefont {X.}~\bibnamefont {Guo}},
  \bibinfo {author} {\bibfnamefont {C.~R.}\ \bibnamefont {Breum}}, \bibinfo
  {author} {\bibfnamefont {J.~S.}\ \bibnamefont {Neergaard-Nielsen}}, \ and\
  \bibinfo {author} {\bibfnamefont {U.~L.}\ \bibnamefont {Andersen}},\ }\href
  {\doibase 10.1126/science.aay4354} {\bibfield  {journal} {\bibinfo  {journal}
  {Science}\ }\textbf {\bibinfo {volume} {366}},\ \bibinfo {pages} {369}
  (\bibinfo {year} {2019})}\BibitemShut {NoStop}%
\bibitem [{\citenamefont {Eaton}\ \emph {et~al.}(2019)\citenamefont {Eaton},
  \citenamefont {Nehra},\ and\ \citenamefont {Pfister}}]{Eaton2019}%
  \BibitemOpen
  \bibfield  {author} {\bibinfo {author} {\bibfnamefont {M.}~\bibnamefont
  {Eaton}}, \bibinfo {author} {\bibfnamefont {R.}~\bibnamefont {Nehra}}, \ and\
  \bibinfo {author} {\bibfnamefont {O.}~\bibnamefont {Pfister}},\ }\href
  {http://arxiv.org/abs/1903.01925} {\bibfield  {journal} {\bibinfo  {journal}
  {arXiv preprint arXiv:1903.01925}\ } (\bibinfo {year} {2019})}\BibitemShut
  {NoStop}%
\bibitem [{\citenamefont {Su}\ \emph {et~al.}(2019)\citenamefont {Su},
  \citenamefont {Myers},\ and\ \citenamefont {Sabapathy}}]{Su2019}%
  \BibitemOpen
  \bibfield  {author} {\bibinfo {author} {\bibfnamefont {D.}~\bibnamefont
  {Su}}, \bibinfo {author} {\bibfnamefont {C.~R.}\ \bibnamefont {Myers}}, \
  and\ \bibinfo {author} {\bibfnamefont {K.~K.}\ \bibnamefont {Sabapathy}},\
  }\href {http://arxiv.org/abs/1902.02331} {\bibfield  {journal} {\bibinfo
  {journal} {arXiv preprint arXiv:1902.02331}\ } (\bibinfo {year}
  {2019})}\BibitemShut {NoStop}%
\bibitem [{\citenamefont {Vasconcelos}\ \emph {et~al.}(2010)\citenamefont
  {Vasconcelos}, \citenamefont {Sanz},\ and\ \citenamefont
  {Glancy}}]{vasconcelos2010all}%
  \BibitemOpen
  \bibfield  {author} {\bibinfo {author} {\bibfnamefont {H.~M.}\ \bibnamefont
  {Vasconcelos}}, \bibinfo {author} {\bibfnamefont {L.}~\bibnamefont {Sanz}}, \
  and\ \bibinfo {author} {\bibfnamefont {S.}~\bibnamefont {Glancy}},\
  }\href@noop {} {\bibfield  {journal} {\bibinfo  {journal} {Optics Letters}\
  }\textbf {\bibinfo {volume} {35}},\ \bibinfo {pages} {3261} (\bibinfo {year}
  {2010})}\BibitemShut {NoStop}%
\bibitem [{\citenamefont {Weigand}\ and\ \citenamefont
  {Terhal}(2018)}]{Weigand2018}%
  \BibitemOpen
  \bibfield  {author} {\bibinfo {author} {\bibfnamefont {D.~J.}\ \bibnamefont
  {Weigand}}\ and\ \bibinfo {author} {\bibfnamefont {B.~M.}\ \bibnamefont
  {Terhal}},\ }\href {\doibase 10.1103/PhysRevA.97.022341} {\bibfield
  {journal} {\bibinfo  {journal} {Physical Review A}\ }\textbf {\bibinfo
  {volume} {97}},\ \bibinfo {pages} {022341} (\bibinfo {year}
  {2018})}\BibitemShut {NoStop}%
\bibitem [{\citenamefont {Dakna}\ \emph
  {et~al.}(1997{\natexlab{a}})\citenamefont {Dakna}, \citenamefont {Anhut},
  \citenamefont {Opatrn{\'{y}}}, \citenamefont {Kn{\"{o}}ll},\ and\
  \citenamefont {Welsch}}]{Dakna1997}%
  \BibitemOpen
  \bibfield  {author} {\bibinfo {author} {\bibfnamefont {M.}~\bibnamefont
  {Dakna}}, \bibinfo {author} {\bibfnamefont {T.}~\bibnamefont {Anhut}},
  \bibinfo {author} {\bibfnamefont {T.}~\bibnamefont {Opatrn{\'{y}}}}, \bibinfo
  {author} {\bibfnamefont {L.}~\bibnamefont {Kn{\"{o}}ll}}, \ and\ \bibinfo
  {author} {\bibfnamefont {D.-G.}\ \bibnamefont {Welsch}},\ }\href {\doibase
  10.1103/PhysRevA.55.3184} {\bibfield  {journal} {\bibinfo  {journal}
  {Physical Review A}\ }\textbf {\bibinfo {volume} {55}},\ \bibinfo {pages}
  {3184} (\bibinfo {year} {1997}{\natexlab{a}})}\BibitemShut {NoStop}%
\bibitem [{\citenamefont {Ourjoumtsev}\ \emph {et~al.}(2006)\citenamefont
  {Ourjoumtsev}, \citenamefont {Tualle-Brouri}, \citenamefont {Laurat},\ and\
  \citenamefont {Grangier}}]{Ourjoumtsev2006}%
  \BibitemOpen
  \bibfield  {author} {\bibinfo {author} {\bibfnamefont {A.}~\bibnamefont
  {Ourjoumtsev}}, \bibinfo {author} {\bibfnamefont {R.}~\bibnamefont
  {Tualle-Brouri}}, \bibinfo {author} {\bibfnamefont {J.}~\bibnamefont
  {Laurat}}, \ and\ \bibinfo {author} {\bibfnamefont {P.}~\bibnamefont
  {Grangier}},\ }\href {\doibase 10.1016/b978-0-408-01434-2.50020-6} {\bibfield
   {journal} {\bibinfo  {journal} {Science}\ }\textbf {\bibinfo {volume}
  {312}},\ \bibinfo {pages} {83} (\bibinfo {year} {2006})}\BibitemShut
  {NoStop}%
\bibitem [{\citenamefont {Neergaard-Nielsen}\ \emph {et~al.}(2006)\citenamefont
  {Neergaard-Nielsen}, \citenamefont {Nielsen}, \citenamefont {Hettich},
  \citenamefont {M{\o}lmer},\ and\ \citenamefont
  {Polzik}}]{Neergaard-Nielsen2006}%
  \BibitemOpen
  \bibfield  {author} {\bibinfo {author} {\bibfnamefont {J.~S.}\ \bibnamefont
  {Neergaard-Nielsen}}, \bibinfo {author} {\bibfnamefont {B.~M.}\ \bibnamefont
  {Nielsen}}, \bibinfo {author} {\bibfnamefont {C.}~\bibnamefont {Hettich}},
  \bibinfo {author} {\bibfnamefont {K.}~\bibnamefont {M{\o}lmer}}, \ and\
  \bibinfo {author} {\bibfnamefont {E.~S.}\ \bibnamefont {Polzik}},\ }\href
  {\doibase 10.1103/PhysRevLett.97.083604} {\bibfield  {journal} {\bibinfo
  {journal} {Physical Review Letters}\ }\textbf {\bibinfo {volume} {97}},\
  \bibinfo {pages} {083604} (\bibinfo {year} {2006})}\BibitemShut {NoStop}%
\bibitem [{\citenamefont {Wakui}\ \emph {et~al.}(2007)\citenamefont {Wakui},
  \citenamefont {Takahashi}, \citenamefont {Furusawa},\ and\ \citenamefont
  {Sasaki}}]{Wakui2007}%
  \BibitemOpen
  \bibfield  {author} {\bibinfo {author} {\bibfnamefont {K.}~\bibnamefont
  {Wakui}}, \bibinfo {author} {\bibfnamefont {H.}~\bibnamefont {Takahashi}},
  \bibinfo {author} {\bibfnamefont {A.}~\bibnamefont {Furusawa}}, \ and\
  \bibinfo {author} {\bibfnamefont {M.}~\bibnamefont {Sasaki}},\ }\href@noop {}
  {\bibfield  {journal} {\bibinfo  {journal} {Optics Express}\ }\textbf
  {\bibinfo {volume} {15}},\ \bibinfo {pages} {3568} (\bibinfo {year}
  {2007})}\BibitemShut {NoStop}%
\bibitem [{\citenamefont {Gerrits}\ \emph {et~al.}(2010)\citenamefont
  {Gerrits}, \citenamefont {Glancy}, \citenamefont {Clement}, \citenamefont
  {Calkins}, \citenamefont {Lita}, \citenamefont {Miller}, \citenamefont
  {Migdall}, \citenamefont {Nam}, \citenamefont {Mirin},\ and\ \citenamefont
  {Knill}}]{Gerrits2010}%
  \BibitemOpen
  \bibfield  {author} {\bibinfo {author} {\bibfnamefont {T.}~\bibnamefont
  {Gerrits}}, \bibinfo {author} {\bibfnamefont {S.}~\bibnamefont {Glancy}},
  \bibinfo {author} {\bibfnamefont {T.}~\bibnamefont {Clement}}, \bibinfo
  {author} {\bibfnamefont {B.}~\bibnamefont {Calkins}}, \bibinfo {author}
  {\bibfnamefont {A.}~\bibnamefont {Lita}}, \bibinfo {author} {\bibfnamefont
  {A.}~\bibnamefont {Miller}}, \bibinfo {author} {\bibfnamefont
  {A.}~\bibnamefont {Migdall}}, \bibinfo {author} {\bibfnamefont {S.~W.}\
  \bibnamefont {Nam}}, \bibinfo {author} {\bibfnamefont {R.}~\bibnamefont
  {Mirin}}, \ and\ \bibinfo {author} {\bibfnamefont {E.}~\bibnamefont
  {Knill}},\ }\href {\doibase 10.1103/PhysRevA.82.031802} {\bibfield  {journal}
  {\bibinfo  {journal} {Physical Review A}\ }\textbf {\bibinfo {volume} {82}},\
  \bibinfo {pages} {031802(R)} (\bibinfo {year} {2010})}\BibitemShut {NoStop}%
\bibitem [{\citenamefont {Hacker}\ \emph {et~al.}(2019)\citenamefont {Hacker},
  \citenamefont {Welte}, \citenamefont {Daiss}, \citenamefont {Shaukat},
  \citenamefont {Ritter}, \citenamefont {Li},\ and\ \citenamefont
  {Rempe}}]{Hacker2019}%
  \BibitemOpen
  \bibfield  {author} {\bibinfo {author} {\bibfnamefont {B.}~\bibnamefont
  {Hacker}}, \bibinfo {author} {\bibfnamefont {S.}~\bibnamefont {Welte}},
  \bibinfo {author} {\bibfnamefont {S.}~\bibnamefont {Daiss}}, \bibinfo
  {author} {\bibfnamefont {A.}~\bibnamefont {Shaukat}}, \bibinfo {author}
  {\bibfnamefont {S.}~\bibnamefont {Ritter}}, \bibinfo {author} {\bibfnamefont
  {L.}~\bibnamefont {Li}}, \ and\ \bibinfo {author} {\bibfnamefont
  {G.}~\bibnamefont {Rempe}},\ }\href {\doibase 10.1038/s41566-018-0339-5}
  {\bibfield  {journal} {\bibinfo  {journal} {Nature Photonics}\ }\textbf
  {\bibinfo {volume} {13}},\ \bibinfo {pages} {110} (\bibinfo {year}
  {2019})}\BibitemShut {NoStop}%
\bibitem [{\citenamefont {Thekkadath}\ \emph {et~al.}(2019)\citenamefont
  {Thekkadath}, \citenamefont {Bell}, \citenamefont {Walmsley},\ and\
  \citenamefont {Lvovsky}}]{thekkadath2019engineering}%
  \BibitemOpen
  \bibfield  {author} {\bibinfo {author} {\bibfnamefont {G.}~\bibnamefont
  {Thekkadath}}, \bibinfo {author} {\bibfnamefont {B.}~\bibnamefont {Bell}},
  \bibinfo {author} {\bibfnamefont {I.}~\bibnamefont {Walmsley}}, \ and\
  \bibinfo {author} {\bibfnamefont {A.}~\bibnamefont {Lvovsky}},\ }\href@noop
  {} {\bibfield  {journal} {\bibinfo  {journal} {arXiv preprint
  arXiv:1908.10314}\ } (\bibinfo {year} {2019})}\BibitemShut {NoStop}%
\bibitem [{\citenamefont {Lita}\ \emph {et~al.}(2008)\citenamefont {Lita},
  \citenamefont {Miller},\ and\ \citenamefont {Nam}}]{Lita2008}%
  \BibitemOpen
  \bibfield  {author} {\bibinfo {author} {\bibfnamefont {A.~E.}\ \bibnamefont
  {Lita}}, \bibinfo {author} {\bibfnamefont {A.~J.}\ \bibnamefont {Miller}}, \
  and\ \bibinfo {author} {\bibfnamefont {S.~W.}\ \bibnamefont {Nam}},\
  }\href@noop {} {\bibfield  {journal} {\bibinfo  {journal} {Optics Express}\
  }\textbf {\bibinfo {volume} {16}},\ \bibinfo {pages} {3032} (\bibinfo {year}
  {2008})}\BibitemShut {NoStop}%
\bibitem [{\citenamefont {Humphreys}\ \emph {et~al.}(2015)\citenamefont
  {Humphreys}, \citenamefont {Metcalf}, \citenamefont {Gerrits}, \citenamefont
  {Hiemstra}, \citenamefont {Lita}, \citenamefont {Nunn}, \citenamefont {Nam},
  \citenamefont {Datta}, \citenamefont {Kolthammer},\ and\ \citenamefont
  {Walmsley}}]{Humphreys2015}%
  \BibitemOpen
  \bibfield  {author} {\bibinfo {author} {\bibfnamefont {P.~C.}\ \bibnamefont
  {Humphreys}}, \bibinfo {author} {\bibfnamefont {B.~J.}\ \bibnamefont
  {Metcalf}}, \bibinfo {author} {\bibfnamefont {T.}~\bibnamefont {Gerrits}},
  \bibinfo {author} {\bibfnamefont {T.}~\bibnamefont {Hiemstra}}, \bibinfo
  {author} {\bibfnamefont {A.~E.}\ \bibnamefont {Lita}}, \bibinfo {author}
  {\bibfnamefont {J.}~\bibnamefont {Nunn}}, \bibinfo {author} {\bibfnamefont
  {S.~W.}\ \bibnamefont {Nam}}, \bibinfo {author} {\bibfnamefont
  {A.}~\bibnamefont {Datta}}, \bibinfo {author} {\bibfnamefont {W.~S.}\
  \bibnamefont {Kolthammer}}, \ and\ \bibinfo {author} {\bibfnamefont {I.~A.}\
  \bibnamefont {Walmsley}},\ }\href@noop {} {\bibfield  {journal} {\bibinfo
  {journal} {New Journal of Physics}\ }\textbf {\bibinfo {volume} {17}},\
  \bibinfo {pages} {103044} (\bibinfo {year} {2015})}\BibitemShut {NoStop}%
\bibitem [{\citenamefont {Zurek}(2001)}]{zurek2001sub}%
  \BibitemOpen
  \bibfield  {author} {\bibinfo {author} {\bibfnamefont {W.~H.}\ \bibnamefont
  {Zurek}},\ }\href {https://www.nature.com/articles/35089017} {\bibfield
  {journal} {\bibinfo  {journal} {Nature}\ }\textbf {\bibinfo {volume} {412}},\
  \bibinfo {pages} {712} (\bibinfo {year} {2001})}\BibitemShut {NoStop}%
\bibitem [{\citenamefont {Duivenvoorden}\ \emph {et~al.}(2017)\citenamefont
  {Duivenvoorden}, \citenamefont {Terhal},\ and\ \citenamefont
  {Weigand}}]{duivenvoorden2017single}%
  \BibitemOpen
  \bibfield  {author} {\bibinfo {author} {\bibfnamefont {K.}~\bibnamefont
  {Duivenvoorden}}, \bibinfo {author} {\bibfnamefont {B.~M.}\ \bibnamefont
  {Terhal}}, \ and\ \bibinfo {author} {\bibfnamefont {D.}~\bibnamefont
  {Weigand}},\ }\href@noop {} {\bibfield  {journal} {\bibinfo  {journal}
  {Physical Review A}\ }\textbf {\bibinfo {volume} {95}},\ \bibinfo {pages}
  {012305} (\bibinfo {year} {2017})}\BibitemShut {NoStop}%
\bibitem [{\citenamefont {Achilles}\ \emph {et~al.}(2004)\citenamefont
  {Achilles}, \citenamefont {Silberhorn}, \citenamefont {Sliwa}, \citenamefont
  {Banaszek}, \citenamefont {Walmsley}, \citenamefont {Fitch}, \citenamefont
  {Jacobs}, \citenamefont {Pittman},\ and\ \citenamefont
  {Franson}}]{achilles2004photon}%
  \BibitemOpen
  \bibfield  {author} {\bibinfo {author} {\bibfnamefont {D.}~\bibnamefont
  {Achilles}}, \bibinfo {author} {\bibfnamefont {C.}~\bibnamefont
  {Silberhorn}}, \bibinfo {author} {\bibfnamefont {C.}~\bibnamefont {Sliwa}},
  \bibinfo {author} {\bibfnamefont {K.}~\bibnamefont {Banaszek}}, \bibinfo
  {author} {\bibfnamefont {I.~A.}\ \bibnamefont {Walmsley}}, \bibinfo {author}
  {\bibfnamefont {M.~J.}\ \bibnamefont {Fitch}}, \bibinfo {author}
  {\bibfnamefont {B.~C.}\ \bibnamefont {Jacobs}}, \bibinfo {author}
  {\bibfnamefont {T.~B.}\ \bibnamefont {Pittman}}, \ and\ \bibinfo {author}
  {\bibfnamefont {J.~D.}\ \bibnamefont {Franson}},\ }\href
  {https://www.tandfonline.com/doi/abs/10.1080/09500340408235288} {\bibfield
  {journal} {\bibinfo  {journal} {Journal of Modern Optics}\ }\textbf {\bibinfo
  {volume} {51}},\ \bibinfo {pages} {1499} (\bibinfo {year}
  {2004})}\BibitemShut {NoStop}%
\bibitem [{\citenamefont {Dakna}\ \emph
  {et~al.}(1997{\natexlab{b}})\citenamefont {Dakna}, \citenamefont {Anhut},
  \citenamefont {Opatrn{\`y}}, \citenamefont {Kn{\"o}ll},\ and\ \citenamefont
  {Welsch}}]{dakna1997generating}%
  \BibitemOpen
  \bibfield  {author} {\bibinfo {author} {\bibfnamefont {M.}~\bibnamefont
  {Dakna}}, \bibinfo {author} {\bibfnamefont {T.}~\bibnamefont {Anhut}},
  \bibinfo {author} {\bibfnamefont {T.}~\bibnamefont {Opatrn{\`y}}}, \bibinfo
  {author} {\bibfnamefont {L.}~\bibnamefont {Kn{\"o}ll}}, \ and\ \bibinfo
  {author} {\bibfnamefont {D.-G.}\ \bibnamefont {Welsch}},\ }\href@noop {}
  {\bibfield  {journal} {\bibinfo  {journal} {Physical Review A}\ }\textbf
  {\bibinfo {volume} {55}},\ \bibinfo {pages} {3184} (\bibinfo {year}
  {1997}{\natexlab{b}})}\BibitemShut {NoStop}%
\end{thebibliography}%


%Control: production of eprint (-1) disabled
%

\end{document}